\documentclass[twocolumn]{aastex62}
\usepackage{amsmath}
\usepackage{amssymb}
\usepackage{graphicx}

\newcommand{\mytilde}{\raise.17ex\hbox{$\scriptstyle\mathtt{\sim}$}}
\newcommand{\Kepler}{\textit{Kepler }}
\newcommand{\Ktwo}{\textit{K2 }}
\newcommand{\TESS}{\textit{TESS }}

\received{}
\revised{}
\accepted{}
\submitjournal{ApJL}

\shorttitle{The 20-20 gap}
\shortauthors{Armstrong, D. J. et al.}

\begin{document}

\title{A Gap in the Mass Distribution for Warm Neptune and Terrestrial Planets}

\email{d.j.armstrong@warwick.ac.uk}

\author[0000-0002-5080-4117]{David J. Armstrong}
\altaffiliation{STFC Ernest Rutherford Fellow}
\affil{Centre for Exoplanets and Habitability, University of Warwick, Gibbet Hill Road, Coventry, CV4 7AL, UK}
\affil{University of Warwick, Department of Physics, Gibbet Hill Road, Coventry, CV4 7AL, UK}

\author{Farzana Meru}
\altaffiliation{Royal Society Dorothy Hodgkin Fellow}
\affil{Centre for Exoplanets and Habitability, University of Warwick, Gibbet Hill Road, Coventry, CV4 7AL, UK}
\affil{University of Warwick, Department of Physics, Gibbet Hill Road, Coventry, CV4 7AL, UK}

\author{Daniel Bayliss}
\affil{Centre for Exoplanets and Habitability, University of Warwick, Gibbet Hill Road, Coventry, CV4 7AL, UK}
\affil{University of Warwick, Department of Physics, Gibbet Hill Road, Coventry, CV4 7AL, UK}

\author{Grant M. Kennedy}
\altaffiliation{Royal Society University Research Fellow}
\affil{Centre for Exoplanets and Habitability, University of Warwick, Gibbet Hill Road, Coventry, CV4 7AL, UK}
\affil{University of Warwick, Department of Physics, Gibbet Hill Road, Coventry, CV4 7AL, UK}

\author{Dimitri Veras}
\altaffiliation{STFC Ernest Rutherford Fellow}
\affil{Centre for Exoplanets and Habitability, University of Warwick, Gibbet Hill Road, Coventry, CV4 7AL, UK}
\affil{University of Warwick, Department of Physics, Gibbet Hill Road, Coventry, CV4 7AL, UK}

\begin{abstract}
Structure in the planet distribution provides an insight into the processes that shape the formation and evolution of planets. The \Kepler mission has led to an abundance of statistical discoveries in regards to planetary radius, but the number of observed planets with measured masses is much smaller. By incorporating results from recent mass determination programs, we have discovered a new gap emerging in the planet population for sub-Neptune mass planets with orbital periods less than 20 days. The gap follows a slope of decreasing mass with increasing orbital period, has a width of a few $M_\oplus$, and is potentially completely devoid of planets. Fitting gaussian mixture models to the planet population in this region favours a bimodel distribution over a unimodel one with a reduction in Bayesian Information Criterion (BIC) of 19.9, highlighting the gap significance. We discuss several processes which could generate such a feature in the planet distribution, including a pileup of planets above the gap region, tidal interactions with the host star, dynamical interactions with the disk, with other planets, or with accreting material during the formation process.
\end{abstract}

\keywords{planets and satellites: general --- planets and satellites: detection --- planets and satellites: dynamical evolution and stability --- planets and satellites: formation --- planets and satellites: physical evolution}

\section{Introduction}

\begin{figure*}
\resizebox{\hsize}{!}{\includegraphics{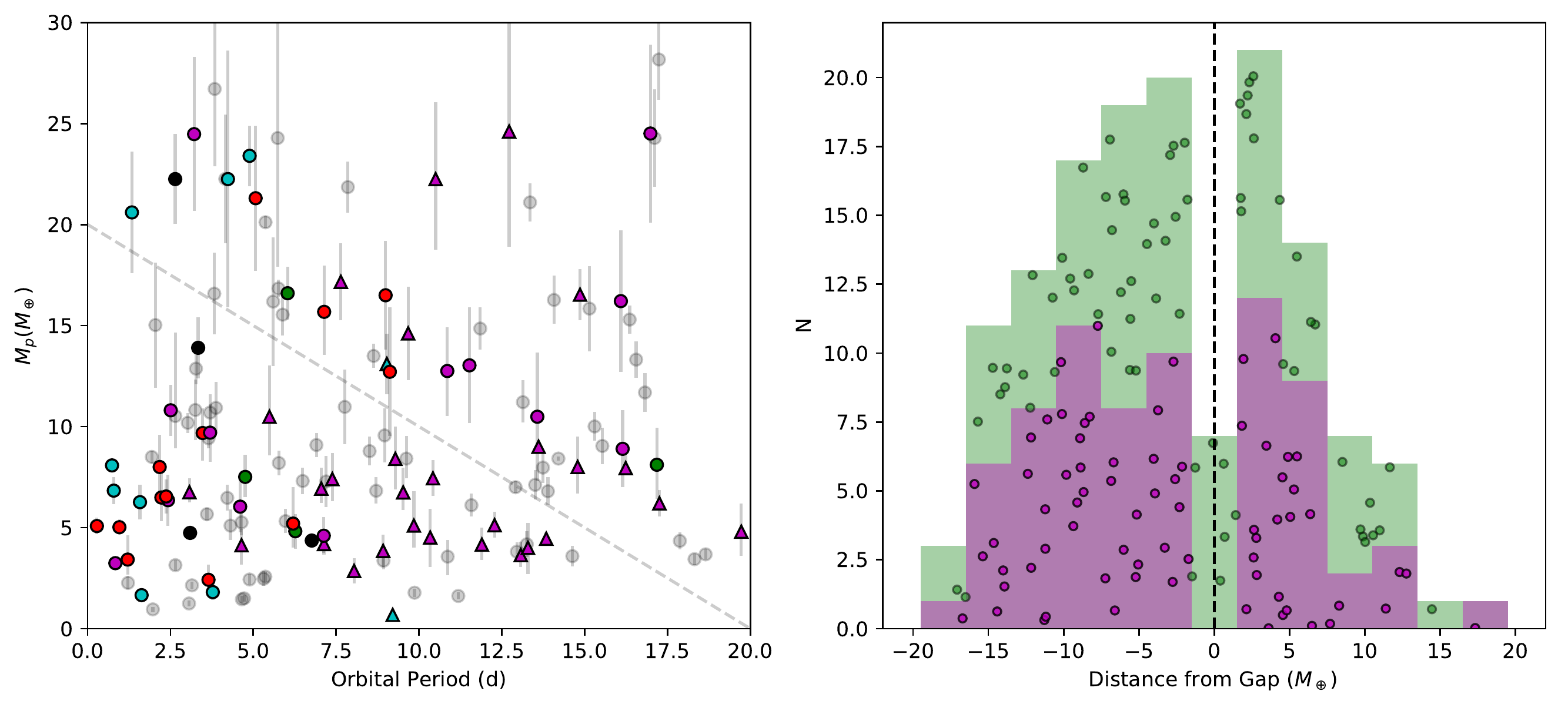}}
\caption{\textbf{Left:} Planets with measured $M_p$ or $M_p\sin(i)$, using $\sin(i)=\pi/4$. Planets with measured inclination (P1 sample) are coloured according to discovery method: Purple - \Kepler, Red - \Ktwo, Green - \TESS, Black - radial velocity surveys, Cyan - mostly ground based photometric surveys. Circular points denote radial velocity derived masses, triangular points masses from transit timing variations. Points without measured inclination (P2 sample) are grey. The gap is shown by a dashed line. \textbf{Right:} Histogram for our P2 (green) and P1 (purple) samples taken on the gradient of the dashed line, limited to planets within $20M_\oplus$. Bins are $3M_\oplus$ wide. Datapoints are shown with a random vertical distribution within the histogram bars.}
\label{figmpplot}
\end{figure*}

Many processes combine to produce the planets we observe, from core formation and accretion at the beginning of a planet's life, to tidal circularisation and orbital decay at the end. With the stream of planet discoveries brought in over the past two decades it has become possible to search for traces of these formation and evolution processes and so place observational limits on their action and effects \citep{Winn:2018er}. One way forward is to study the distribution of planets as a whole, both in terms of planetary and host star parameters. Signatures of their history may remain in the planet population, providing a pathway to testing population synthesis models \citep[e.g.][]{Mordasini:2015ds}.

Several such signatures can be found in the planet radius-period plane, such as the Neptunian desert \citep{Mazeh:2016dz,Owen:2018dh} and `radius gap' \citep{Fulton:2017bp} likely arising from photoevaporation. On a broader scale, the observed occurrence rate of planets has been studied in detail by the \Kepler mission, leading to the discovery that Neptune and Earth size planets are much more common than those of Jupiter size, as well as providing increasingly improved estimates of the frequency of Earth-like planets. \citep[e.g.][]{Fressin:2013df,Mulders:2018ex,Hsu:2019ur}.

In the planet mass-period plane we can expect to view the history of planets from a different angle. While photoevaporation can significantly change a planet's radius, the effect on planetary mass is predicted to be more modest, at least for orbital periods larger than a few days \citep{Owen:2017kf,Jin:2018ef}. Current studies of the planet mass population allow investigations of the effects of tides on giant planets \citep{Bonomo:2017fl} and the occurrence rate of planets out to 20 AU \citep{Bryan:2016cl}. A similar Neptunian desert is seen, thought to arise from a combination of tidal interactions with the host star and high eccentricity migration \citep{Matsakos:2016id,Owen:2018dh}.

Here we present a new emerging signature in the planet mass-orbital period plane, a gap for planets with mass less than \mytilde 20 $M_\oplus$ and period less than $20$d. The physical reasons for the gap remain unexplained and are left for future exploration, though we provide some plausible hypotheses. These may provide important constrains on planet formation, migration and star-planet tidal theory.

\section{Planet Sample}
We use the confirmed planet sample from the NASA Exoplanet Archive\footnote{https://exoplanetarchive.ipac.caltech.edu/}, as of the 24th May 2019. Our prime sample (hereafter P1) consists of all transiting planets with measured masses $M_p$, radii $R_p$ and hence inclination, within the limits $M_p < 25M_\oplus$ and orbital period $P < 20$ days. These are all transiting planets with masses determined through radial velocities (RVs) or transit timing variations (TTVs). Our second wider sample (hereafter P2) consists of all planets with $M_p$ or $M_p\sin(i)$ within the same limits and additionally contains planets with no measured inclination. Only planets with mass measurements better than $3\sigma$ were included in either sample. For plots and calculations involving $M_p\sin(i)$, we assumed $\sin(i) = \pi/4$, the average for an isotropic distribution.

We manually check each archive entry for the P1 sample to ensure accuracy. For Kepler-266d and e the values in the archive did not match the best fit values from the source publication \citep{Rodriguez:2018cx} and were updated, for Kepler-10b we used values from the more recent \citet{Rajpaul:2017ef} as they were inconsistent with the default catalogue values, and KOI-142b was removed from the sample as the given source publication \citep{Nesvorny:2013cb} did not contain a mass measurement for the inner planet. In several cases a more recent publication was available but with results consistent with the archive default value. In these cases, we used the archive default for consistency.

After these steps, our P1 sample contained 72 planets, and our P2 sample contained 143 planets including the P1 sample. The full sample, parameters used and source references are given in Table \ref{tabsample}.

\section{The M-P Gap}
We plot our planet sample and histogram in Figure \ref{figmpplot}. A gap in the distribution is seen following a straight line gradient of \mytilde$-1M_\oplus d^{-1}$, with a width of a few $M_\oplus$. The gap can be seen in both the P2 and P1 samples, although it is clearer for P1, potentially due to the blurring effect of unknown inclination. Below we discuss the relevant observational biases, and conduct tests to determine the significance of the apparent gap.

\subsection{Observational biases}
As we are using a planet sample drawn from a variety of discovery and characterisation programs the biases underlying the sample are complex. We do not attempt to calculate planetary occurrence rates, although we note the \TESS satellite should provide a sample of planets more amenable to such a study by the end of its planned mission \citep{Barclay:2018hn}. 

All planets are subject to a bias on detection and on characterisation. Our P1 sample (72 total) contains planets discovered by the \Kepler mission (41), \Ktwo mission (13), \TESS mission (4), various other photometric surveys (10) and RV surveys (4). The remaining planets that complete the P2 sample are all discovered through RVs. Full details are in Table \ref{tabsample}. Figure \ref{figmpplot} shows the various discovery sources in relation to the gap. We note that points both above and below the gap arise from a mix of different discovery missions and methods, indicating that these factors do not cause the observed gap. The gap remains significant if ground-based planet discoveries are removed (Section \ref{sectHartigan}).

For planets with known radii, which in our sample were all measured by the transit method, there is a bias against low $(R_p/R_*)^2$ due to decreasing signal strength, and large semi-major axis $a$ as the probability of transit decreases with $1/a$ and is limited by mission baseline. Most transiting planets in the sample were discovered via the Kepler mission, for which the detection efficiency falls for planets with periods longer than several hundred days, and radii smaller than \mytilde 1--2$R_\oplus$ for FGK dwarf stars \citep{Thompson:2018gm}, well outside the range covered by the gap. For ground based photometric surveys and to a lesser extent K2 the period and radius sensitivity is more relevant, with ground based surveys in particular struggling to detect planets of Earth and Neptune size aside from a handful of cases with M dwarf host stars. For planets with RV based detection or characterisation, there is a bias in the strength of the Doppler signal arising from the stellar reflex motion, which goes as $M_p\sin(i)M_*^{-2/3}P^{-1/3}$. From both transits and RVs there is then a bias against planets with long periods and low masses or radii. The hypothesised gap gradient has a different sign to this trend, and hence is caused by a different mechanism.

Our sample contains planets with masses measured through both RVs and TTVs. A key concern is whether the use of two methods is creating the appearance of two populations in the data. For the TTV characterised planets, mass measurements are only possible in multiple planetary systems. There is a further bias towards systems in or near mean motion resonances, as well as to larger planet masses, and larger, transiting, companion planets which are perturbed by the mass of the planet under question. We note that the majority of our TTV-derived masses are from the \Kepler mission, where the data precision and short period of these systems lead to reliable mass determinations. For a full discussion of the nature of the TTV signal and its potential biases we refer the reader to the extensive body of literature \citep{Hadden:2014bf,Steffen:2016ke,Mills:2017hp,Agol:2018gj}. Figure \ref{figmpplot} shows the source of the mass measurement for each point, and Section \ref{sectHartigan} demonstrates that the gap arises in both the RV and TTV populations. 

Additionally stellar activity, host star brightness, available telescope time and its effects on observing cadence, plus the choice by various teams of particular planets to target for mass measurement, are all potential biases. Although these effects might lead to trends in the $M_p$-$P$ plane, we would not expect any of them to preferentially avoid planets in a few $M_\oplus$ wide region near the line in Figure \ref{figmpplot}. The host star brightness is investigated in Section \ref{sectDiscuss}, and is a proxy for the likelihood of a target to be selected for characterisation, as brighter host stars require shorter exposure times and are more scientifically interesting for further follow-up. The gap shows no dependence on host star brightness.

\begin{table*}
\caption{Planet Sample. Full table available online.}
\label{tabsample}
\begin{tabular}{llllllllllr}
\hline
\hline
Planet & Sample & P & $M_p$ &  $R_p$ & $\rho_p$ & $e_p$ & $M_*$ & Multiplicity & Facility & Reference \\
 & & d & $M_\oplus$ &  $R_\oplus$ & $gcm^{-3}$ &  & $M_\odot$ &  &  &  \\
\hline
K2-291 b & P1 & 2.2252 & $6.490^{+1.160}_{-1.160}$ &  1.59 & 8.84 & 0.0 & 0.93 & 1 & Kepler & \citet{Kosiarek:2019gi} \\
K2-265 b & P1 & 2.3692 & $6.541^{+0.839}_{-0.839}$ &  1.72 & 7.1 & 0.084 & 0.92 & 1 & K2 & \citet{Lam:2018ez} \\
HD 51608 b & P2 & 14.073 & $12.78^{+1.208}_{-1.176}$ & nan & nan & 0.09 & 0.8 & 2 & La Silla & \citet{Udry:2019cr} \\
\vdots  & \vdots & \vdots & \vdots & \vdots & \vdots & \vdots& \vdots & \vdots & \vdots \\ 
\hline
\end{tabular}
\end{table*}

\subsection{Gap Width and Gradient}
\label{sectgapgrad}
Given the observational biases involved it is premature to characterise the width or depth of the gap in detail. We do however estimate the width and gradient here to help inform a theoretical understanding of the processes that could lead to such a gap. We use the P1 sample as the two populations are more clearly separated in Figure \ref{figmpplot}.

To estimate the gap gradient and width we employ a support vector machine \citep{Cortes:1995ie} with a linear kernel, applied using the scikit-learn support vector classification tool \citep{Pedregosa:2011tv}. A simple support vector machine finds the decision boundary separating two classes such that the classes are separated by as clean and wide a gap as possible. In our case we train the classifier by assigning points above the dashed line in Figure \ref{figmpplot} to one class and those below to the other. After fitting, the resulting boundary has a gradient of $-0.90M_\oplus d^{-1}$ and offset of $18.9M_\oplus$. The smallest gap between two points in the P1 sample using this gradient is $4.0M_\oplus$, an estimate of the gap width. Note that this analysis makes no conclusion as to the gap significance, which is explored below.

\subsection{Hartigan's Dip Test}
\label{sectHartigan}
\begin{figure}
\resizebox{\hsize}{!}{\includegraphics{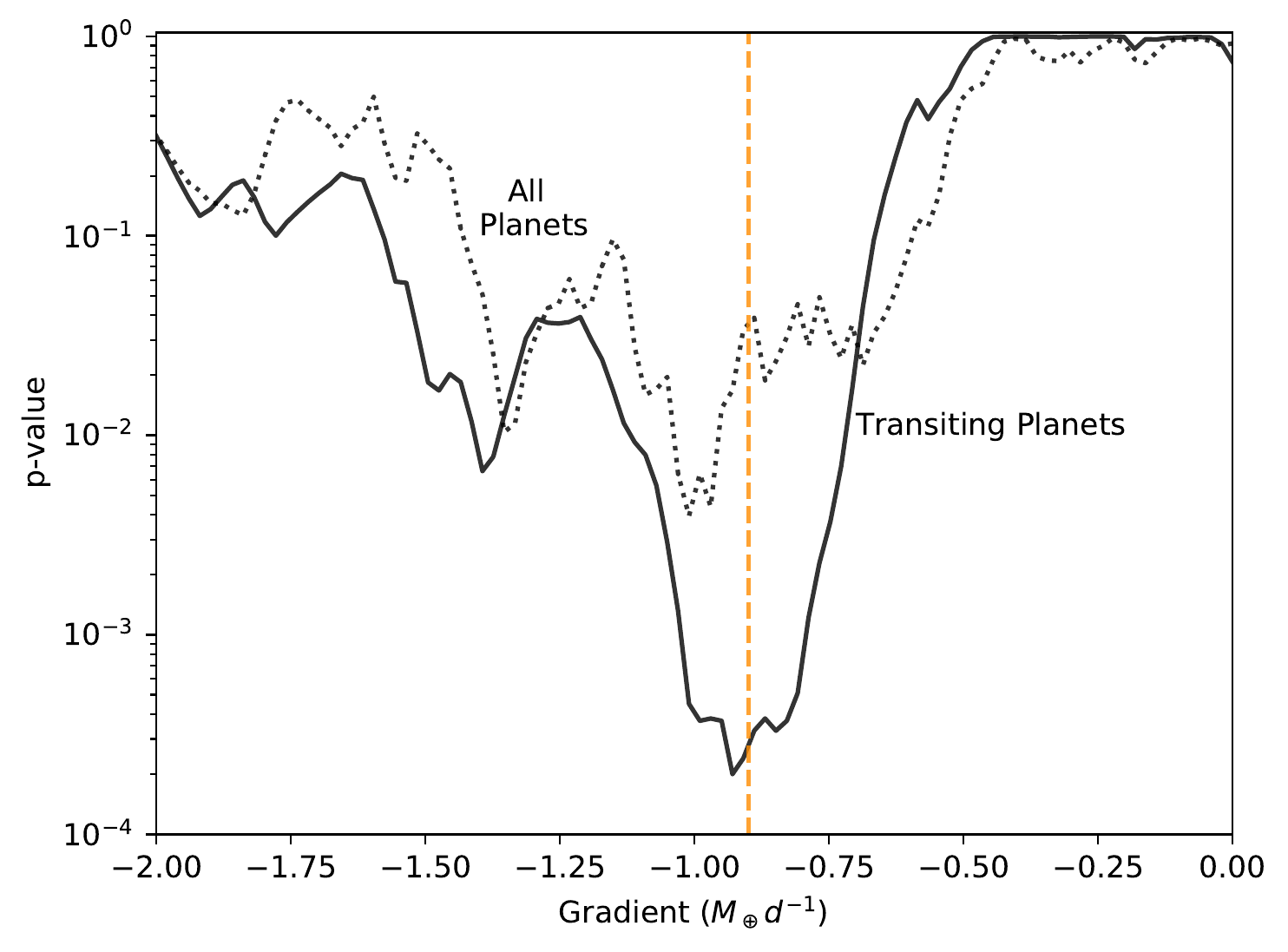}}
\caption{p-value output as a function of gap gradient from a bootstrapped Hartigan dip test applied to our P1 sample (solid line) and P2 sample (dotted line). The vertical dashed line marks the gradient found in Section \ref{sectgapgrad}.}
\label{fighartigan}
\end{figure}

Hartigan's dip test \citep{Hartigan:1985cu} is a test of multimodality in a sample population, which compares the empirical sample distribution to a generalised unimodal distribution. We apply the dip test to our P1 and P2 samples in Figure \ref{fighartigan}. We trial a range of gap gradients between 0 and -2$M_\oplus d^{-1}$, rotating the data into a frame parallel to the tested gradient each time, as the dip test operates in one dimension. We measure the significance of the results by generating 100000 trials of uniformly distributed random data covering the same parameter space, rotated through the same angle, for each tested gradient.

To measure significance, we fix the gradient to $-0.90M_\oplus d^{-1}$, the value found in Section \ref{sectgapgrad}. For the P1 sample we find a p-value of $2.4 \times 10^{-4}$, and for the P2 $3.6 \times 10^{-2}$. We note that the gradient was extracted from the P1 sample, and so the gradient choice here is not optimal for the P2 sample, as seen in Figure \ref{fighartigan}.

As a test of the underlying biases, we extract the gap significance at the same gradient for several subsamples of the P1 set. The first consists of only planets detected by space-based observatories, i.e. \Kepler, \Ktwo and \TESS (58 planets). The p-value found is $9.2 \times 10^{-4}$, demonstrating that any bias affecting ground-based surveys, such as window functions or airmass-induced systematics, does not generate the gap. We then consider planets with masses derived from TTVs only (30 planets), giving a p-value of $2.2 \times 10^{-2}$, and planets with masses derived from RVs (42 planets), giving a p-value of $3.2 \times 10^{-4}$. In all cases the gap remains significant.

\subsection{Gaussian Mixture Models}

As another test of gap significance we consider Gaussian Mixture Models (GMM). We implement GMMs using the scikit-learn package \citep{Pedregosa:2011tv}, with no priors on the component Gaussian locations or covariances. We incorporate measurement errors on the data by using 10000 bootstrap samples, sampling each time from the normal distribution defined by each datapoint's value and error. For each trial, we fit independent GMMs with between one and five components. Each GMM is refit 100 times with random initial parameters, and the best fitting result taken and stored. For each trial we measure the BIC of the best fitting model.

\begin{figure}
\resizebox{\hsize}{!}{\includegraphics{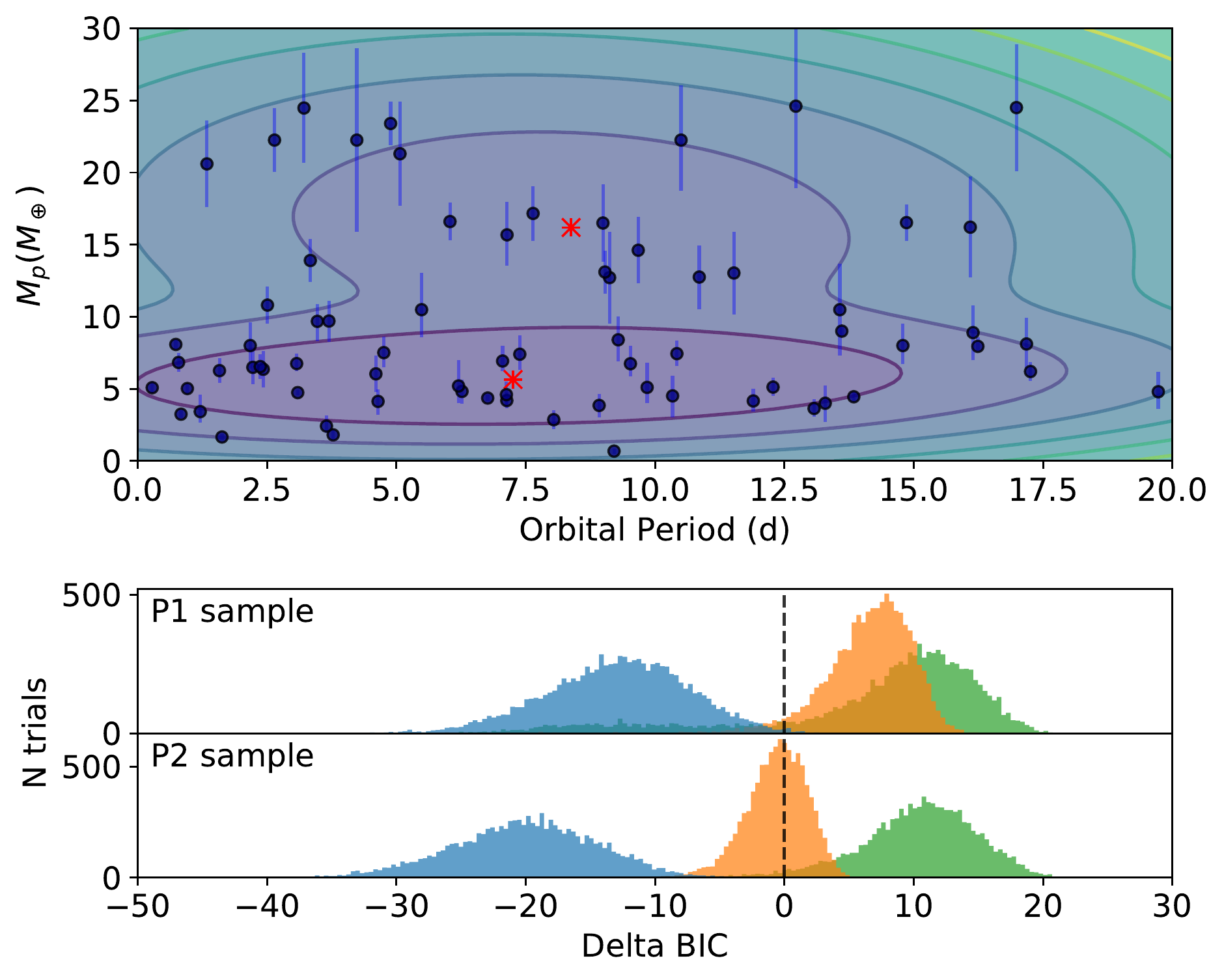}}
\caption{\textbf{Top:} Two-component GMM fit to the P1 sample with contours showing weighted log probabilities of the combined model. Contour levels are from 2.5 to 8.5 with a spacing of 1 in log probability. The red crosses mark the component means. \textbf{Bottom:} Distribution of $\Delta$BIC over 10000 bootstrap trials when varying the number of components. Blue: $1 \rightarrow 2$, Orange: $2 \rightarrow 3$ and Green: $3 \rightarrow 4$.}
\label{figGMMall}
\end{figure}

We consider the above test performed on the P2 and P1 samples. For both samples the results show strong support for a two-component model, with a change in BIC of $-19.9^{+5.5}_{-5.8}$ over the 10000 samples when moving from one to two components in the P2 sample, with none of the draws resulting in increased BIC. For the P1 sample the two-component model is also most favoured with a change in BIC of $-12.7^{+5.2}_{-5.8}$, with 59 out of 10000 draws giving increased BIC. Three, four, and five component models are rejected as the BIC increases each time. Figure \ref{figGMMall} shows the distribution of changes in BIC (lower panel) and a contour of the weighted log probabilities of an example fitted two-component model (top panel).

Curiously the results are strongest for the P2 sample despite the two populations being more well separated in the P1 sample. This is due to the increased number of datapoints; this increase in significance with more data, despite the extra blurring factor introduced by unknown inclination, supports the physical reality of the gap.

\section{Discussion}
\label{sectDiscuss}
\begin{figure*}
\resizebox{\hsize}{!}{\includegraphics{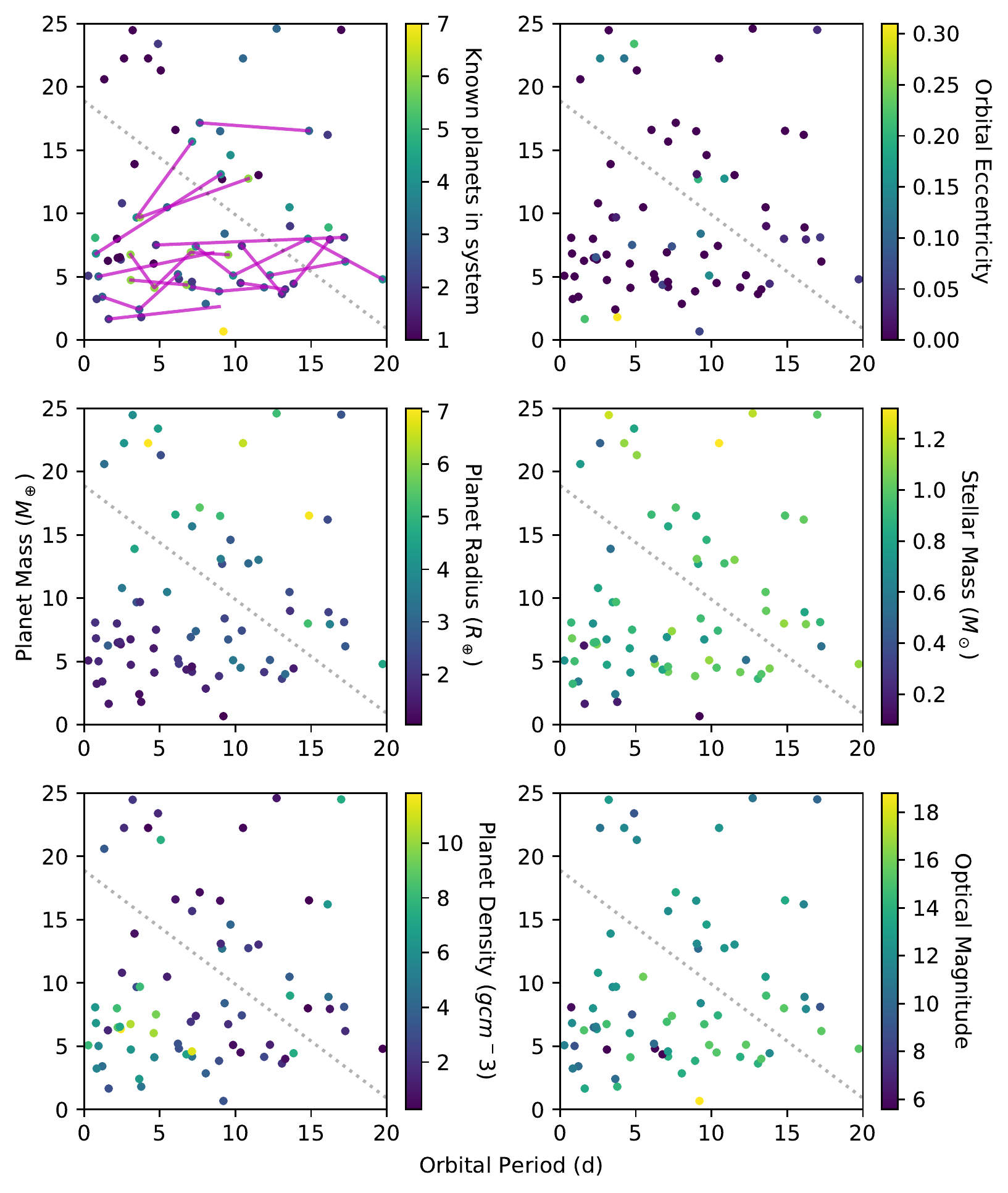}}
\caption{The P1 sample coloured by several potentially important parameters. From top left: Known planets in system with multiples joined by solid magenta lines, planetary eccentricity, planet radius, host star mass, planet density and optical magnitude of the host star.}
\label{figdepen}
\end{figure*}

Figure \ref{figdepen} shows the dependency of the gap on various stellar and planetary parameters for ease of reference. No clear trends are seen. In particular, we note that the mass-period gap occurs at higher planetary masses to the photoevaporation valley of \citet{Fulton:2017bp}. While the photoevaporation valley is poorly constrained at these small semi-major axes, the valley would have to rise significantly as well as be associated with more mass-loss than previously thought to lead to the gap observed here. We discuss a range of potential hypotheses and their likelihood to contribute to the gap formation below. 

\subsection{An accretion/ejection boundary?}

The fate of small bodies interacting with a planet depends on the ability of that planet to accrete or eject said bodies. As discussed by \citet{Wyatt:2016fq} this balance is set by the ratio of the Keplerian velocity $v_{\rm Kep}$ from the star at the planet's semi-major axis, and the escape velocity from the planet $v_{\rm esc}$. Expressed in terms of orbital period, this ratio is
\begin{equation}\label{eq:esc}
    M_{\rm pl} = 16000 M_\star P_{\rm pl}^{-1} \rho_{\rm pl}^{-1/2} \left(v_{\rm esc}/v_{\rm Kep}\right)^3 \, M_\oplus ,
\end{equation}
where the units of $M_\star$ is Solar masses, period is in days, and density is in g\,cm$^-3$. Thus, the accretion/ejection criterion gives approximately the correct dependence of planet mass with period over the range where data exist, but with a normalisation that is approximately 160$\times$ too high if the boundary lies in the gap.

How might the accretion/ejection boundary apply here? There is no obvious answer, but the idea would need to be based on local growth. Perhaps the boundary does lie in the gap due to inefficient growth, and planets grow locally towards it but no higher. However, some of the most massive planets reach even higher masses by late pairwise collisions \citep[e.g.][]{Izidoro:2017jr}, which approximately double their masses, thus leaving a gap just above the boundary.

\subsection{Multi-planet stability}
\label{sectpileup}
An alternative view of the observations is as a pileup of planets above the gap. If nearly every system contained a planet in the pileup region, dynamical instability could lead to the emergence of a gap in the distribution. For multi-planet systems, \cite{Gladman1993} showed that a planetary system can only be stable if the orbital separation between two neighbouring planets is larger than a critical separation, implying empty regions on either side of a hypothetical pileup.

While planet stability could in principle describe the existence of such a gap, another explanation is needed to explain why a pileup would form in the appropriate region of parameter space, and occur so uniformly across planetary systems. We give one potential pileup formation mechanism in the next section. Furthermore, this hypothesis would not explain why the seemingly single-planet systems are also not located in the gap, unless those systems contain as yet unseen planets.

\subsection{Zero-torque location}

In young protoplanetary discs it is possible that planets become trapped in a `€œzero-torque location'€" i.e. a radial location where the sum of the torques acting on the planet cancel out such that the planet does not migrate. Such a scenario could explain the pileup of Section \ref{sectpileup}. These zero torque locations move radially inwards as the disc evolves over time \citep{Lyra_etal2010}.  It can be seen from \citet{Bitsch:2015gt} and \citet{Morbidelli:2016im} that at late times in the evolution of a disc the zero torque boundary follows a curve where essentially the planet mass falls off with radial location, as seen in our observational results, though the slope of this boundary is yet to be analysed in detail and the simulations are limited to outer regions of the disc. Further study of other planet-disk processes including gap-opening and migration is needed to fully assess the importance of the disc phase.

\subsection{Tides}
The radial extent of the gap coincides entirely with the region where star-planet tides would affect both the spin and orbital evolution of planets. Hence, regardless of where the planet is formed and when and how it migrated, a planet entering into this region would effectively ``activate" tides. Consequently, we speculate that the gap may be a natural result of spin-orbit tidal interactions, for which current understanding may be limited.

For example, as shown by \cite{Efroimsky:2013hj}, the constant phase lag model \citep{Goldreich:1966jo}, while convenient, is physically and mathematically inconsistent. Further, the constant time lag model \citep{Hut:1981vc,Eggleton:1998ch} has very limited applicability \citep{Makarov:2015in}. Capture into spin-orbit resonances and psuedo-synchronous rotation states cannot be accurately realised with such models \citep{Makarov:2013iv,Noyelles:2014jn}. Instead, the direction of the net tidally-induced migration is chaotic with respect to the multi-dimensional parameter space \citep{Veras:2019}, and hence very well might produce gaps like the one seen here.

\section{Conclusion}
We have demonstrated the significance of a gap in the planet distribution for close-in Neptune and terrestrial mass planets. In the planet mass -- orbital period plane, the gap cuts an approximately straight line from ($20 M_{\oplus}$, $0$ days) to ($0 M_{\oplus}$, $20$ days). Several mechanisms could be responsible for the gap formation, including tidal interactions with the host star, dynamical interactions with the disk, with other planets, or with accreting material. New discoveries from the \TESS satellite among others will reveal this feature in more detail.

\section*{Acknowledgements}
We thank the anonymous referee for a thorough review which improved the paper. This paper includes data collected by the Kepler and TESS missions, which are publicly available from the Mikulski Archive for Space Telescopes (MAST). Funding for the Kepler and TESS missions is provided by NASA'€™s Science Mission directorate. This research has made use of the NASA Exoplanet Archive, which is operated by the California Institute of Technology, under contract with the National Aeronautics and Space Administration under the Exoplanet Exploration Program. DJA and DV respectively acknowledge support from the STFC via Ernest Rutherford Fellowships (ST/R00384X/1) and (ST/P003850/1). GMK is supported by the Royal Society as a Royal Society University Research Fellow. F.M. acknowledges support from the Royal Society Dorothy Hodgkin Fellowship.

\bibliography{papers_140319}

\begin{thebibliography}{}
\expandafter\ifx\csname natexlab\endcsname\relax\def\natexlab#1{#1}\fi
\providecommand{\url}[1]{\href{#1}{#1}}
\providecommand{\dodoi}[1]{doi:~\href{http://doi.org/#1}{\nolinkurl{#1}}}
\providecommand{\doeprint}[1]{\href{http://ascl.net/#1}{\nolinkurl{http://ascl.net/#1}}}
\providecommand{\doarXiv}[1]{\href{https://arxiv.org/abs/#1}{\nolinkurl{https://arxiv.org/abs/#1}}}

\bibitem[{Agol \& Fabrycky(2018)}]{Agol:2018gj}
Agol, E., \& Fabrycky, D.~C. 2018, in Handbook of Exoplanets (Cham: Springer,
  Cham), 797--816

\bibitem[{Barclay {et~al.}(2018)Barclay, Pepper, \& Quintana}]{Barclay:2018hn}
Barclay, T., Pepper, J., \& Quintana, E.~V. 2018, The Astrophysical Journal
  Supplement Series, 239, 2

\bibitem[{Bitsch {et~al.}(2015)Bitsch, Johansen, Lambrechts, \&
  Morbidelli}]{Bitsch:2015gt}
Bitsch, B., Johansen, A., Lambrechts, M., \& Morbidelli, A. 2015, Astronomy and
  Astrophysics, 575, A28

\bibitem[{Bonomo {et~al.}(2017)Bonomo, Desidera, Benatti, Borsa, Crespi,
  Damasso, Lanza, Sozzetti, Lodato, Marzari, Boccato, Claudi, Cosentino,
  Covino, Gratton, Maggio, Micela, Molinari, Pagano, Piotto, Poretti,
  Smareglia, Affer, Biazzo, Bignamini, Esposito, Giacobbe, H{\'e}brard,
  Malavolta, Maldonado, Mancini, Fiorenzano, Masiero, Nascimbeni, Pedani,
  Rainer, \& Scandariato}]{Bonomo:2017fl}
Bonomo, A.~S., Desidera, S., Benatti, S., {et~al.} 2017, Astronomy and
  Astrophysics, 602, A107

\bibitem[{Bryan {et~al.}(2016)Bryan, Knutson, Howard, Ngo, Batygin, Crepp,
  Fulton, Hinkley, Isaacson, Johnson, Marcy, \& Wright}]{Bryan:2016cl}
Bryan, M.~L., Knutson, H.~A., Howard, A.~W., {et~al.} 2016, The Astrophysical
  Journal, 821, 89

\bibitem[{Cortes \& Vapnik(1995)}]{Cortes:1995ie}
Cortes, C., \& Vapnik, V. 1995, Machine Learning, 20, 273

\bibitem[{Efroimsky \& Makarov(2013)}]{Efroimsky:2013hj}
Efroimsky, M., \& Makarov, V.~V. 2013, The Astrophysical Journal, 764, 26

\bibitem[{Eggleton {et~al.}(1998)Eggleton, Kiseleva, \& Hut}]{Eggleton:1998ch}
Eggleton, P.~P., Kiseleva, L.~G., \& Hut, P. 1998, The Astrophysical Journal,
  499, 853

\bibitem[{Fressin {et~al.}(2013)Fressin, Torres, Charbonneau, Bryson,
  Christiansen, Dressing, Jenkins, Walkowicz, \& Batalha}]{Fressin:2013df}
Fressin, F., Torres, G., Charbonneau, D., {et~al.} 2013, The Astrophysical
  Journal, 766, 81

\bibitem[{Fulton {et~al.}(2017)Fulton, Petigura, Howard, Isaacson, Marcy,
  Cargile, Hebb, Weiss, Johnson, Morton, Sinukoff, Crossfield, \&
  Hirsch}]{Fulton:2017bp}
Fulton, B.~J., Petigura, E.~A., Howard, A.~W., {et~al.} 2017, The Astronomical
  Journal, 154, 109

\bibitem[{{Gladman}(1993)}]{Gladman1993}
{Gladman}, B. 1993, \icarus, 106, 247, \dodoi{10.1006/icar.1993.1169}

\bibitem[{Goldreich(1966)}]{Goldreich:1966jo}
Goldreich, P. 1966, The Astronomical Journal, 71, 1

\bibitem[{Hadden \& Lithwick(2014)}]{Hadden:2014bf}
Hadden, S., \& Lithwick, Y. 2014, The Astrophysical Journal, 787, 80

\bibitem[{Hartigan \& Hartigan(1985)}]{Hartigan:1985cu}
Hartigan, J.~A., \& Hartigan, P.~M. 1985, The Annals of Statistics, 13, 70

\bibitem[{Hsu {et~al.}(2019)Hsu, Ford, Ragozzine, \& Ashby}]{Hsu:2019ur}
Hsu, D.~C., Ford, E.~B., Ragozzine, D., \& Ashby, K. 2019

\bibitem[{Hut(1981)}]{Hut:1981vc}
Hut, P. 1981, Astronomy and Astrophysics, 99, 126

\bibitem[{Izidoro {et~al.}(2017)Izidoro, Ogihara, Raymond, Morbidelli, Pierens,
  Bitsch, Cossou, \& Hersant}]{Izidoro:2017jr}
Izidoro, A., Ogihara, M., Raymond, S.~N., {et~al.} 2017, Monthly Notices of the
  Royal Astronomical Society, 470, 1750

\bibitem[{Jin \& Mordasini(2018)}]{Jin:2018ef}
Jin, S., \& Mordasini, C. 2018, The Astrophysical Journal, 853, 163

\bibitem[{Kosiarek {et~al.}(2019)Kosiarek, Blunt, Lopez-Morales, Crossfield,
  Sinukoff, Petigura, Gonzales, Poretti, Malavolta, Howard, Isaacson, Haywood,
  Ciardi, Bristow, Cameron, Charbonneau, Dressing, Figueira, Fulton, Hardee,
  Hirsch, Latham, Mortier, Nava, Schlieder, Vanderburg, Weiss, Bonomo, Bouchy,
  Buchhave, Coffinet, Damasso, Dumusque, Lovis, Mayor, Micela, Molinari, Pepe,
  Phillips, Piotto, Rice, Sasselov, Segransan, Sozzetti, Udry, \&
  Watson}]{Kosiarek:2019gi}
Kosiarek, M.~R., Blunt, S., Lopez-Morales, M., {et~al.} 2019, The Astronomical
  Journal, 157, 116

\bibitem[{Lam {et~al.}(2018)Lam, Santerne, Sousa, Vigan, Armstrong, Barros,
  Brugger, Adibekyan, Almenara, Mena, Dumusque, Barrado, Bayliss, Bonomo,
  Bouchy, Brown, Ciardi, Deleuil, Demangeon, Faedi, Foxell, Jackman, King,
  Kirk, Ligi, Lillo-Box, Lopez, Lovis, Louden, Nielsen, McCormac, Mousis,
  Osborn, Pollacco, Santos, Udry, \& Wheatley}]{Lam:2018ez}
Lam, K. W.~F., Santerne, A., Sousa, S.~G., {et~al.} 2018, Astronomy and
  Astrophysics, 620, A77

\bibitem[{{Lyra} {et~al.}(2010){Lyra}, {Paardekooper}, \& {Mac
  Low}}]{Lyra_etal2010}
{Lyra}, W., {Paardekooper}, S.-J., \& {Mac Low}, M.-M. 2010, \apj, 715, L68,
  \dodoi{10.1088/2041-8205/715/2/L68}

\bibitem[{Makarov(2015)}]{Makarov:2015in}
Makarov, V.~V. 2015, The Astrophysical Journal, 810, 12

\bibitem[{Makarov \& Efroimsky(2013)}]{Makarov:2013iv}
Makarov, V.~V., \& Efroimsky, M. 2013, The Astrophysical Journal, 764, 27

\bibitem[{Matsakos \& K{\"o}nigl(2016)}]{Matsakos:2016id}
Matsakos, T., \& K{\"o}nigl, A. 2016, The Astrophysical Journal Letters, 820,
  L8

\bibitem[{Mazeh {et~al.}(2016)Mazeh, Holczer, \& Faigler}]{Mazeh:2016dz}
Mazeh, T., Holczer, T., \& Faigler, S. 2016, Astronomy and Astrophysics, 589,
  A75

\bibitem[{Mills \& Mazeh(2017)}]{Mills:2017hp}
Mills, S.~M., \& Mazeh, T. 2017, The Astrophysical Journal Letters, 839, L8

\bibitem[{Morbidelli \& Raymond(2016)}]{Morbidelli:2016im}
Morbidelli, A., \& Raymond, S.~N. 2016, Journal of Geophysical Research
  (Planets), 121, 1962

\bibitem[{Mordasini {et~al.}(2015)Mordasini, Molli{\`e}re, Dittkrist, Jin, \&
  Alibert}]{Mordasini:2015ds}
Mordasini, C., Molli{\`e}re, P., Dittkrist, K.~M., Jin, S., \& Alibert, Y.
  2015, International Journal of Astrobiology, 14, 201

\bibitem[{Mulders {et~al.}(2018)Mulders, Pascucci, Apai, \&
  Ciesla}]{Mulders:2018ex}
Mulders, G.~D., Pascucci, I., Apai, D., \& Ciesla, F.~J. 2018, The Astronomical
  Journal, 156, 24

\bibitem[{Nesvorn{\'{y}} {et~al.}(2013)Nesvorn{\'{y}}, Kipping, Terrell,
  Hartman, Bakos, \& Buchhave}]{Nesvorny:2013cb}
Nesvorn{\'{y}}, D., Kipping, D., Terrell, D., {et~al.} 2013, The Astrophysical
  Journal, 777, 3

\bibitem[{Noyelles {et~al.}(2014)Noyelles, Frouard, Makarov, \&
  Efroimsky}]{Noyelles:2014jn}
Noyelles, B., Frouard, J., Makarov, V.~V., \& Efroimsky, M. 2014, Icarus, 241,
  26

\bibitem[{Owen \& Lai(2018)}]{Owen:2018dh}
Owen, J.~E., \& Lai, D. 2018, Monthly Notices of the Royal Astronomical
  Society, 479, 5012

\bibitem[{Owen \& Wu(2017)}]{Owen:2017kf}
Owen, J.~E., \& Wu, Y. 2017, The Astrophysical Journal, 847, 29

\bibitem[{Pedregosa {et~al.}(2011)Pedregosa, Varoquaux, Gramfort, Michel,
  Thirion, Grisel, Blondel, Prettenhofer, Weiss, Dubourg, Vanderplas, Passos,
  Cournapeau, Brucher, Perrot, \& Duchesnay}]{Pedregosa:2011tv}
Pedregosa, F., Varoquaux, G., Gramfort, A., {et~al.} 2011, Journal of Machine
  Learning Research, 12, 2825

\bibitem[{Rajpaul {et~al.}(2017)Rajpaul, Buchhave, \& Aigrain}]{Rajpaul:2017ef}
Rajpaul, V., Buchhave, L.~A., \& Aigrain, S. 2017, Monthly Notices of the Royal
  Astronomical Society: Letters, 471, L125

\bibitem[{Rodriguez {et~al.}(2018)Rodriguez, Becker, Eastman, Hadden,
  Vanderburg, Khain, Quinn, Mayo, Dressing, Schlieder, Ciardi, Latham,
  Rappaport, Adams, Berlind, Bieryla, Calkins, Esquerdo, Kristiansen,
  Omohundro, Schwengeler, Stassun, \& Terentev}]{Rodriguez:2018cx}
Rodriguez, J.~E., Becker, J.~C., Eastman, J.~D., {et~al.} 2018, The
  Astronomical Journal, 156, 245

\bibitem[{Steffen(2016)}]{Steffen:2016ke}
Steffen, J.~H. 2016, Monthly Notices of the Royal Astronomical Society, 457,
  4384

\bibitem[{Thompson {et~al.}(2018)Thompson, Coughlin, Hoffman, Mullally,
  Christiansen, Burke, Bryson, Batalha, Haas, Catanzarite, Rowe, Barentsen,
  Caldwell, Clarke, Jenkins, Li, Latham, Lissauer, Mathur, Morris, Seader,
  Smith, Klaus, Twicken, Van~Cleve, Wohler, Akeson, Ciardi, Cochran, Henze,
  Howell, Huber, Prsa, Ramirez, Morton, Barclay, Campbell, Chaplin,
  Charbonneau, Christensen-Dalsgaard, Dotson, Doyle, Dunham, Dupree, Ford,
  Geary, Girouard, Isaacson, Kjeldsen, Quintana, Ragozzine, Shabram, Shporer,
  Aguirre, Steffen, Still, Tenenbaum, Welsh, Wolfgang, Zamudio, Koch, \&
  Borucki}]{Thompson:2018gm}
Thompson, S.~E., Coughlin, J.~L., Hoffman, K., {et~al.} 2018, The Astrophysical
  Journal Supplement Series, 235, 38

\bibitem[{Udry {et~al.}(2019)Udry, Dumusque, Lovis, S{\'e}gransan, D{\'\i}az,
  Benz, Bouchy, Coffinet, Lo~Curto, Mayor, Mordasini, Motalebi, Pepe, Queloz,
  Santos, Wyttenbach, Alonso, Cameron, Deleuil, Figueira, Gillon, Moutou,
  Pollacco, \& Pompei}]{Udry:2019cr}
Udry, S., Dumusque, X., Lovis, C., {et~al.} 2019, Astronomy and Astrophysics,
  622, A37

\bibitem[{{Veras} {et~al.}(2019){Veras}, {Efroimsky}, {Makarov}, {Bou{\'e}},
  {Wolthoff}, {Reffert}, {Quirrenbach}, {Tremblay}, \&
  {G{\"a}nsicke}}]{Veras:2019}
{Veras}, D., {Efroimsky}, M., {Makarov}, V.~V., {et~al.} 2019, \mnras, 486,
  3831, \dodoi{10.1093/mnras/stz965}

\bibitem[{Winn(2018)}]{Winn:2018er}
Winn, J.~N. 2018, in Handbook of Exoplanets (Cham: Springer, Cham), 1--18

\bibitem[{Wyatt {et~al.}(2016)Wyatt, Bonsor, Jackson, Marino, \&
  Shannon}]{Wyatt:2016fq}
Wyatt, M.~C., Bonsor, A., Jackson, A.~P., Marino, S., \& Shannon, A. 2016,
  Monthly Notices of the Royal Astronomical Society, 464, 3385

\end{thebibliography}
\bibliographystyle{aasjournal}

\end{document}